\documentclass[a4paper,aps,11pt,preprint,showpacs]{revtex4}
\usepackage{graphicx,amsmath,revsymb,bm,amssymb,amsfonts,amsthm}
\begin{document}

\title{Excitation and photoionization processes involving the
bound $ns$ electrons}

\author{A.V.~Nefiodov$^{\,\mathrm{a,b,}}$}
\thanks{Corresponding author.  \\
\textit{\hphantom{*}E-mail address}:
nefiodov@theory.phy.tu-dresden.de (A.V. Nefiodov).}
\author{G.~Plunien$^{\,\mathrm{a}}$}
\affiliation{$^{\mathrm{a}}$Institut f\"ur Theoretische Physik,
Technische Universit\"at Dresden, Mommsenstra{\ss}e 13, D-01062
Dresden, Germany \\
$^{\mathrm{b}}$Petersburg Nuclear Physics Institute, 188300
Gatchina, St.~Petersburg, Russia}

\date{Received \today}
\widetext
\begin{abstract}
We have considered the processes of excitation and ionization of
light multicharged ions by impact of high-energy particles, which
proceed with participation of the $ns$ electrons. The screening
corrections to the energy levels and photoionization cross sections
are evaluated analytically within the framework of the
non-relativistic perturbation theory with respect to the
electron-electron interaction. The universal scalings for the
excitation and ionization cross sections are studied for arbitrary
principal quantum numbers $n$.
\end{abstract}
\maketitle

{\bf 1.} Since decades the fundamental processes of excitation and
ionization of few-electron atomic ions have been persistently
investigated within the framework of different sophisticated
approaches, due to necessity of the accurate account of all
interactions in the colliding system (see, for example, the works
\cite{1,2,3,4,5,6,7} and references there). The deduction of the
universal scaling behavior for differential and total cross sections
is of particular importance, because it allows one to establish
generic features of various processes for a wide family of targets
\cite{8,9,10,11}. In this Letter, we study the excitation and
photoionization of light multicharged ions, which proceed with
participation of the $ns$ electrons. As a method, the consistent
non-relativistic perturbation theory in the Furry picture is
employed \cite{12}. The calculations are performed analytically,
taking into account the one-photon exchange diagrams.

The characteristic quantities for the theoretical description of
collision processes on multicharged ions are the Coulomb potential
$I=\eta^2/(2m)$ for single ionization from the K shell, the average
mo\-men\-tum $\eta=m \alpha Z$ of the K-shell electron, the Bohr
radius $a_0=1/(m\alpha)$, the electron mass $m$, and the
fine-structure constant $\alpha$ ($\hbar=1$, $c=1$). The parameter
$\alpha Z$ is supposed to be sufficiently small ($\alpha Z \ll 1$),
although we assume nuclear charges with $Z \gg 1$.

{\bf 2.} In the non-relativistic theory, the stationary states of
hydrogen-like atomic system are characterized by the principal
quantum number $n$, the value of angular momentum $l$, and
projection of the orbital angular momentum $m$ \cite{13}. The
corresponding eigenfunctions, which are solutions of the
Schr{\"o}dinger equation for a bound electron in the external
Coulomb field of a point nucleus, read \cite{14,15}
\begin{eqnarray}
\psi_{nlm}(\bm{r}) &=& R_{nl}(r) Y_{lm}(\theta,\varphi) \, ,
\label{eq1} \\
R_{nl}(r)&=& - 2 \eta_n^{3/2}\frac{\sqrt{(n-l-1)!}}{\sqrt{[(n+l)!]^3
n}}\, e^{- \eta_n r} (2 \eta_n r)^{l} L_{n+l}^{2l+1}(2 \eta_n r)
\,  , \label{eq2} \\
E_{nl}&=& - I_n= - \frac{I}{n^2} \, . \label{eq3}
\end{eqnarray}
Here $\eta_n=\eta/n$ and $Y_{lm}$ are the spherical harmonics. The
wave functions are normalized in the standard fashion
\begin{gather}
\int d\bm{r} \, \psi^*_{nlm}(\bm{r}) \psi_{n'l'm'}(\bm{r}) =
\delta_{nn'} \delta_{ll'} \delta_{mm'} \,  , \label{eq4} \\
\int\limits^\infty_0 dr r^2 R_{nl}(r) R_{n'l}(r) = \delta_{nn'}\,  .
\end{gather}
According to works \cite{16,17,18}, it is convenient to represent
the associated Laguerre polynomials in Eq.~\eqref{eq2} via the
contour integral
\begin{equation}
L_{n+l}^{2l+1}(\zeta) = - \frac{(n+l)!}{2 \pi i}
\oint\limits^{(0^+)} \! dt \frac{\exp[-\zeta t/(1-t)]}{(1-t)^{2l+2}
t^{n-l}}  \,  , \label{eq6}
\end{equation}
where the closed path encircles counter-clockwise the origin $t=0$,
but not the point $t=1$.

In the following, we shall focus on the bound $ns$ states ($l=0$).
In the momentum representation, the corresponding eigenfunctions
\eqref{eq1} can be written as
\begin{eqnarray}
\psi_{ns}(\bm{f}) &=& \frac{1}{\sqrt{4 \pi}} \, R_{n0}(f)  \, ,
\label{eq7} \\
R_{n0}(f)&=& \frac{\sqrt{\eta_n}}{n} \frac{1}{2 \pi i}
\oint\limits^{(\eta_n^+)} \! dy \frac{(y+\eta_n)^{n}}{(y -
\eta_n)^{n}}\Bigl(-\frac{\partial}{\partial \lambda} \Bigr) \langle
\bm{f}| V_{i\lambda} | 0 \rangle_{\left|\vphantom{\bigl(}\, \lambda
= y \right. }  \,   ,  \label{eq8} \\
\langle \bm{f}'| V_{i\lambda} | \bm{f}  \rangle &=& \frac{4\pi}{(
\bm{f}'- \bm{f} )^2 + \lambda^2} \,  . \label{eq9}
\end{eqnarray}
In Eq.~\eqref{eq8}, after taking the derivative with respect to
$\lambda$, one should set $\lambda=y$ and then perform the contour
integration enclosing the pole at $y=\eta_n$.

{\bf 3.} Let us consider a helium-like ion in the pure
$ns^2\,({}^1\!S)$ state. Within the framework of non-relativistic
perturbation theory with respect to the electron-electron
interaction, the energy levels $E$ are given as a series in powers
of the reversed nuclear charge $Z^{-1}$
\begin{equation}
E =  E^{(0)} + \sum\limits_{k \geqslant 1} \Delta E^{(k)} \equiv I
\sum\limits_{k \geqslant 0} \epsilon_k Z^{-k}  \,  , \label{eq10}
\end{equation}
where $\epsilon_0 =-2n^{-2}$. The dimensionless coefficients
$\epsilon_k$ depend on the principal quantum number $n$. The
first-order correlation correction $\Delta E^{(1)}$ can be written
as
\begin{equation}
\Delta E^{(1)} = 4 \pi \alpha \int \frac{d\bm{f}}{(2\pi)^{3}}
\frac{d\bm{f}_1}{(2\pi)^{3}}\frac{d\bm{f}_2}{(2\pi)^{3}} \langle
\psi_{ns} |\bm{f}_1\rangle \langle \bm{f}_1+\bm{f}| \psi_{ns}\rangle
\frac{1}{f^2}\langle \psi_{ns} |\bm{f}_2\rangle \langle
\bm{f}_2-\bm{f}| \psi_{ns}\rangle  \,  . \label{eq11}
\end{equation}
Performing integrations over the intermediate momenta yields
\begin{eqnarray}
\Delta E^{(1)} &=& \alpha N_{ns}\frac{1}{2 n^3 } \frac{1}{(2 \pi
i)^3} \oint\limits^{(\eta_n^+)} \! \prod
\limits_{k=1}^{3}\biggl[dy_k
\frac{(y_k+\eta_n)^{n}}{(y_k - \eta_n)^{n}} \biggr]\nonumber\\
&&\times \Bigl(-\frac{\partial}{\partial \lambda} \Bigr)
\frac{1}{\lambda^{2}}\langle \psi_{ns}| \left(V_{iy_3} -
V_{i(\lambda + y_3 )}\right) | 0 \rangle_{\left|\vphantom{\bigl(}\,
\lambda = y_1 + y_2 \right. }\,  , \label{eq12}
\end{eqnarray}
where $N^2_{ns}= \eta_n^3/\pi$ and $\eta_n=\eta/n$. The matrix
element evaluated with the Coulomb wave function \eqref{eq1} reads
\cite{19}
\begin{equation}
\langle \psi_{nlm} |V_{i \lambda} |0 \rangle = 4 \pi N_{ns}
\frac{(\lambda - \eta_n)^{n-1}}{(\lambda + \eta_n)^{n+1}}\,
\delta_{l0} \delta_{m0} \,  .  \label{eq13}
\end{equation}
The integrals in the expression \eqref{eq12} are given by residues
of the integrand at the poles $y_k =\eta_n$, $(k=1,2,3)$. For the
ground state ($n=1$), $\epsilon_1= 5/4$ \cite{15}, while for the
$2s^2$ configuration, $\epsilon_1= 77/256$ \cite{20}. In
Table~\ref{table1}, we present the coefficients $\epsilon_1= Z
\Delta E^{(1)}/I$ calculated for the principal quantum numbers $n
\leqslant 16$. In the asymptotic limit $n \gg 1$, $n^2\epsilon_1$
tends to the constant $1.189$.

As another example, we shall also evaluate the correlation
correction $\Delta E^{(1)}$ for helium-like ion in the
$1sns\,({}^{1,3}\!S)$ states. The energy levels $E$ can be again
presented as an expansion \eqref{eq10}. However, in this case, for
$k \geqslant 1$, the coefficients $\epsilon_k$ differ for the
singlet and triplet states, while $\epsilon_0 =-(1+n^{-2})$. The
first-order correction $\Delta E^{(1)}$ contains contributions of
the Coulomb direct and exchange integrals
\begin{eqnarray}
\Delta E^{(1)} &=& J\pm K \,  ,   \label{eq14}\\
J &=& 2\pi \alpha \,\eta^{-1} N_{1s}^2
N^{\vphantom{2}}_{ns}\frac{1}{2 \pi i} \oint\limits^{(\eta_n^+)} \!
dy \frac{(y+\eta_n)^{n}}{(y - \eta_n)^{n}} \nonumber\\
&&\times \Bigl(-\frac{\partial}{\partial \lambda} \Bigr)
\frac{1}{\lambda^{2}}\langle \psi_{ns}| \left(V_{iy} - V_{i(\lambda
+ y)}\right) | 0 \rangle_{\left|\vphantom{\bigl(}\,
\lambda = 2 \eta \right. }\,  , \label{eq15}\\
K &=& 2\pi \alpha \,\eta^{-1}N_{1s}^2
N^{\vphantom{2}}_{ns}\frac{1}{2 \pi i} \oint\limits^{(\eta_n^+)} \!
dy \frac{(y+\eta_n)^{n}}{(y - \eta_n)^{n}} \nonumber\\
&&\times \Bigl(-\frac{\partial}{\partial \lambda} \Bigr)
\frac{1}{\lambda^{2}}\langle \psi_{ns}| \left(V_{i\eta} -
V_{i(\lambda + \eta)}\right) | 0 \rangle_{\left|\vphantom{\bigl(}\,
\lambda = y + \eta \right. }\,  . \label{eq16}
\end{eqnarray}
In Eq.~\eqref{eq14}, the plus and minus signs correspond to the
singlet and triplet states, respectively. The matrix elements with
the Coulomb wave functions are given by Eq.~\eqref{eq13}. The
splitting between the energy levels with different multiplicities is
just $2K$. In Tables~\ref{table2} and \ref{table3}, we present the
coefficients $\epsilon_1=Z \Delta E^{(1)}/I$ calculated for the
$n\,{}^{1,3}\!S$ terms with $n \leqslant 15$. In the asymptotic
limit $n \gg 1$, both coefficients $\epsilon_1$ exhibit a similar
behavior as $2n^{-2}$, because the contribution due to the exchange
interaction vanishes.

{\bf 4.} Let us now consider the high-energy electron scattering on
a hydrogen-like ion  being in the ground state, which results in
excitation of a K-shell electron into the bound $ns$ state
($n\geqslant 2$). We shall derive formulas for differential and
total cross sections of the process in the leading order of
non-relativistic perturbation theory. The particular case of $n=2$
has been studied in Refs.~\cite{20,21}. The incident electron is
characterized by the energy $E_p=\bm{p}^2/(2 m)$ and the momentum
$\bm{p}$ at infinitely large distances from the nucleus, while the
scattered electron possesses the energy $E_{p_1}=\bm{p}_1^2/(2 m)$
and the asymptotic momentum $\bm{p}_1$. The energy-conservation law
implies $E_p +E_{1s}=E_{p_1} + E_{ns}$.

On the example of the $1s$-$2s$ excitation \cite{21}, we have seen
that the calculations of total cross sections performed within the
framework of the Born approximation are worthwhile even in the
near-threshold domain, where the expansion with respect to the
powers of the reversed energy $E_p^{-1}$ appears to be an asymptotic
series. The best agreement with the exact calculations is achieved,
if one truncates the expansion with taking into account only the
leading high-energy term.

\begin{figure}[h]
\centerline{\includegraphics[scale=0.6]{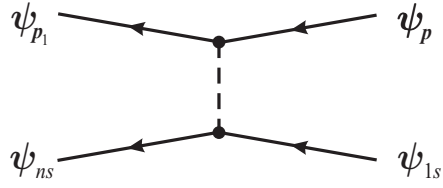}}
\caption{\label{fig1} Feynman diagram for excitation of a K-shell
electron into the bound $ns$ state by electron impact. Solid lines
denote electrons in the external Coulomb field of the nucleus, while
the dashed line denotes the electron-electron Coulomb interaction.}
\end{figure}

For non-relativistic energies $E_p$ within the asymptotic range
$I(1- n^{-2}) \ll E_p \ll m$, $E_{p_1} \sim E_p$ and the absolute
value of the asymptotic momentum of the scattered electron is
estimated as $p_1 \sim p \gg \eta$. Accordingly, one needs to
calculate only the Feynman diagram depicted in Fig.~\ref{fig1}. The
wave functions of both the incident and scattered high-energy
electrons can be approximated by plane waves (Born approximation).
The contribution of the exchange diagram turns out to be
significantly suppressed and thus negligible. Using the expressions
\eqref{eq7}--\eqref{eq9}, the amplitude of the process can be
represented as follows
\begin{eqnarray}
{\mathcal A} &=& 4 \pi \alpha \, N_{1s} \frac{1}{q^2} \Bigl( -
\frac{\partial}{\partial \lambda} \Bigr) \langle
\psi_{{ns}}|V_{i\lambda}| \bm{q} \rangle_{\left|\vphantom{\bigl(}\,
\lambda = \eta\right. } \,  , \label{eq17}\\
\langle \psi_{{ns}}|V_{i\lambda}| \bm{q} \rangle &=& N_{ns}
\frac{\pi}{i \eta q}\biggl\{ \frac{(\nu^* + \eta_{n})^{n}}{(\nu^* -
\eta_{n})^{n}} - \frac{(\nu + \eta_{n})^{n}}{(\nu -
\eta_{n})^{n}}\biggr\}\,  , \label{eq18}
\end{eqnarray}
where $N^2_{ns}=\eta_n^3/\pi$, $\nu= iq - \lambda$, and
$\bm{q}=\bm{p}- \bm{p}_1$ is the momentum transfer. Although
Eq.~\eqref{eq18} is written in the complex form, it is a real
function of the square of the momentum transfer $q^2$. In the limit
$q^2 \to 0$, Eq.~\eqref{eq18} is in agreement with the expression
\eqref{eq13}. Taking the derivative with respect to $\lambda$ in
Eq.~\eqref{eq17} yields
\begin{eqnarray}
{\mathcal A} &=& 4 \pi \alpha \, \eta^{-5} N_{1s}N_{ns} T_n(x)
\,  , \label{eq19}\\
T_n(x) &=& \frac{2 \pi}{i x^3} \biggl\{ \frac{(ix -
\tau_{-})^{n-1}}{(ix - \tau_{+})^{n+1}} - \frac{(ix +
\tau_{-})^{n-1}}{(ix + \tau_{+})^{n+1}}\biggr\}  \,  , \label{eq20}
\end{eqnarray}
where $x=q/\eta$ and $\tau_{\pm}= 1\pm n^{-1}$. Note, that $T_n(x)$
is a real function depending actually on $x^2$.

The differential cross section for the $1s$-$ns$ excitation is
related to the dimensionless function \eqref{eq20} via
\begin{equation}
d\sigma^*_{1s}(ns) =\frac{\sigma_0}{Z^4} \frac{4\,T_n^2(x)}{\pi^2
n^3 \varepsilon}\, dx^2 \,   ,
\label{eq21}                                          
\end{equation}
where $\sigma_0=\pi a_0^2=87.974$ Mb. Here we have also introduced
the dimensionless energy $\varepsilon =E_{p}/I$ of the incident
electron. The energy-conservation law implies $\varepsilon =
\varepsilon_1 + 1- n^{-2}$, where $\varepsilon_1=E_{p_1}/I$ denotes
the dimensionless energy of the scattered electron. The asymptotic
non-relativistic energy domain is characterized by $1-n^{-2} \ll
\varepsilon \ll 2 (\alpha Z)^{-2}$.

The leading high-energy contribution to the total cross section for
the excitation process is given by
\begin{eqnarray}
\sigma^*_{1s}(ns) &=& \frac{\sigma_0}{Z^4} \, Q_n(\varepsilon)
\, ,   \quad  (n\geqslant 2)   \,  , \label{eq22}\\ 
Q_n(\varepsilon)&=& \frac{\varkappa_n}{n^3 \varepsilon} \,   ,
\label{eq23}       \\                               
\varkappa_n &=& \frac{8}{\pi^2} \int\limits_{0}^{\infty} dx \,x
\,T_n^2(x)  \nonumber\\
&=& \frac{2^{6}}{(2n+1)!} \frac{d^{2n+1}}{dx^{2n+1}} \biggl[
\frac{\ln x}{x^{5}}(x - i \tau_{-})^{2n-2}
\biggr]_{\left|\vphantom{\bigl(}\, x=i\tau_{+}\right. } \nonumber\\
&& -\frac{2^{7}}{n!} \frac{d^{n}}{dx^{n}} \biggl[ \frac{\ln
x}{x^{5}} \frac{(x^2 + \tau^2_{-})^{n-1}}{(x + i \tau_{+})^{n+1}}
\biggr]_{\left|\vphantom{\bigl(}\, x=i\tau_{+}\right. } \, .
\label{eq24}                                        
\end{eqnarray}
In Eq.~\eqref{eq24}, we have chosen the regular branch of the
logarithm, which assumes real values on the upper edge of the cut
made along the positive semi-axis. The universal function
$Q_n(\varepsilon)$ does not depend on the nuclear charge $Z$ and
describes the excitation of states with the zeroth-order matrix
element of the dipole transition \cite{15}. The dimensionless
quantity $\varkappa_n$, which is a function of the principal quantum
number $n$, is presented for particular values of $n \leqslant 12$
in Table \ref{table4}. In the limit $n \gg 1$, the coefficients
$\varkappa_n$ tend to the asymptotic value $\varkappa_n \simeq
1.798$.

Let us make a few comments:
\begin{itemize}
\item  Due to the crossing symmetry \cite{12}, the Feynman graph
depicted in Fig.~\ref{fig1} describes also the $1s$-$ns$ excitation
by the high-energy positron impact. The exchange effect is absent at
all. To get the amplitude for the process, one needs to make the
following substitutions: $\bm{p} \rightleftharpoons -\bm{p}_1$,
which do not alter expressions for the ionization cross section.
Accordingly, Eqs.~\eqref{eq22}--\eqref{eq24} are also valid for the
case of the positron impact.

\item Although we have considered the excitation process by the
high-energy electron impact, Eqs.~\eqref{eq22}--\eqref{eq24} are
also valid for the case of fast projectiles with another mass $M$.
The corresponding energy-conservation law still reads $E_p - I =
E_{p_1} - I_n$, where $E_p=M v^2/2$ and $I_n$ is given by
Eq.~\eqref{eq3}. However, now it is convenient to calibrate the
energies of the incident and scattered particle by the
characteristic binding energy $\tilde{I} = M(\alpha Z)^2/2$, namely,
$\varepsilon= E_p/\tilde{I}$ and $\varepsilon_1 =
E_{p_1}/\tilde{I}$. The dimensionless energy
$\varepsilon=v^2/(\alpha Z)^2$ does not depend on the mass of the
incident particle, while the energy-conservation law now implies
$\varepsilon= \varepsilon_1 + \mu(1- n^{-2})$, where $\mu = m/M$.
The asymptotic energy range is characterized by $\mu(1-n^{-2}) \ll
\varepsilon \ll 2(\alpha Z)^{-2}$.

\item  To leading order of non-relativistic perturbation theory,
the total cross section for impact excitation of helium-like ions
from the ground state into the $1s ns$ configuration is by a factor
2 as large as that $\sigma^*_{1s}(ns)$ for hydrogen-like ions,
taking into account the number of target electrons.
\end{itemize}

\begin{figure}[h]
\centerline{\includegraphics[scale=0.6]{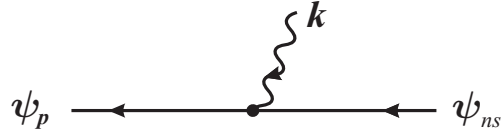}}
\caption{\label{fig2} Feynman diagram for ionization of the bound
$ns$ electron by photon impact.}
\end{figure}

{\bf 5.} The process of the single photoionization of a
hydrogen-like ion in the $ns$ state is described by the diagram
depicted in Fig.~\ref{fig2}. The incident photon is characterized by
the momentum $\bm{k}$, the energy $\omega = |\bm{k}| = k $, and the
polarization vector $\bm{\mathrm e}$. The energy-conservation law
reads $E_p = \omega - I_n$, where $E_p =\bm{p}^2/(2 m)$ is the
energy of the outgoing electron and $I_n$ is given by
Eq.~\eqref{eq3}. Accordingly, the non-relativistic photoeffect can
proceed at photon energies $I_n \leqslant \omega \ll m$. In the
dipole approximation, the non-relativistic problem can be solved
analytically \cite{22,23,24,25,26}. Using the integral
representation \eqref{eq8} for the Coulomb wave functions
\eqref{eq7} yields the amplitude of the process under consideration
in the closed form
\begin{eqnarray}
\mathcal{A}_{ns} &=& 4 \pi \eta^{-3} N_{\gamma} N_{ns}   N_{p} (1 -
i\xi)(\bm{\mathrm{e}}\cdot \bm{p}) \, \Phi_n(\xi)  \,   ,
\label{eq25}\\
\Phi_n(\xi) &=& \frac{1}{(n-1)!} \frac{d^{n-1}}{dy^{n-1}} \biggl[
\frac{y (y+n^{-1})^{n}}{(y^2+ \xi^{-2})^{2}} \, e^{-2 \xi
\cot^{-1}(y\xi)} \biggr]_{\left|\vphantom{\bigl(}\, y=1/n \right.
} \,  , \label{eq26}\\
N_{\gamma} &=& \frac{1}{m} \frac{\sqrt{4 \pi \alpha}}{\sqrt{2
\omega}} \,  , \quad  N^2_{ns}= \frac{\eta_n^3}{\pi}   \,  , \quad
N^2_{p} = \frac{2\pi\xi} {1 - e^{-2\pi \xi}} \,  ,
\end{eqnarray}
where $\xi=\eta/p$. Here we employ the Coulomb gauge, in which
$(\bm{\mathrm e}\cdot \bm{k}) = 0$ and $(\bm{\mathrm e}^*\cdot
\bm{\mathrm e}) = 1$. The dimensionless function \eqref{eq26} can be
written as
\begin{equation}
\Phi_n(\xi)=  \frac{e^{-2 \xi \cot^{-1}(\xi/n)}}{(n^2 + \xi^2)^{n}}
\,\xi^{4} f_n(\xi) \,   ,
\label{eq28}                                          
\end{equation}
where $f_n(\xi)$ for particular values of $1\leqslant n\leqslant 9$
are presented in Table \ref{table5}. For large values of $n$, it is
convenient to employ the recurrence relations between the matrix
elements \cite{22,23,24,25,26}.

The total cross section reads \cite{23,25}
\begin{eqnarray}
\sigma^{+}_{ns} &=&  \frac{\sigma_{0}}{Z^2} F_n(\xi)\,  , \quad
n^{-2} \leqslant \varepsilon_{\gamma} \ll 2(\alpha Z)^{-2} \,  ,
\label{eq29}\\
F_n(\xi)&=& \frac{2^7 \pi}{3 n} \frac{(1+ \xi^{2})\xi^{8} }{(1 -
e^{-2\pi \xi})} \frac{e^{-4 \xi \cot^{-1}(\xi/n)}}{(n^2 +
\xi^2)^{2n+1}} f^2_n(\xi)   \,  , \label{eq30}
\end{eqnarray}
where $\sigma_0=\alpha \pi a_0^2=0.642$ Mb. Due to the
energy-conservation law, the parameter $\xi=\eta/p$ corresponds to
the dimensionless energy of the photon $\varepsilon_{\gamma}=
\omega/I$ according to $\varepsilon_{\gamma}=\xi^{-2} + n^{-2}$. The
universal function $F_n(\xi)$ does not depend on the nuclear charge
number $Z$.

In the high-energy non-relativistic domain, which is characterized
by $n^{-2} \ll \varepsilon_{\gamma} \ll 2(\alpha Z)^{-2} $, the
amplitude \eqref{eq25} simplifies and appears as
\begin{equation}
\mathcal{A}_{ns} = N_{\gamma} N_{ns} \left(\bm{\mathrm e} \cdot
\bm{p} \right) \frac{8 \pi \eta}{q^4}=n^{-3/2}\mathcal{A}_{1s} \,  .
\label{eq31}
\end{equation}
Here $\bm{q}=\bm{p} - \bm{k} \simeq \bm{p}$, because the process
proceeds with a large momentum transfer $q \gg \eta$. The total
cross section reads \cite{14}
\begin{equation}
\sigma^{+}_{ns} =  \frac{\sigma_{0}}{Z^2} \frac{2^8}{3 n^3}\,
\xi^{7} =\frac{\sigma^{+}_{1s}}{n^3} \,  ,  \label{eq32}
\end{equation}
where $\xi \simeq \varepsilon_{\gamma}^{-1/2}$. The formula
\eqref{eq32}, which provides just the leading term in the expansion
of Eq.~\eqref{eq29} with respect to the parameter $\xi \ll 1$, can
be obtained within the Born approximation. Since the function
\eqref{eq30} involves also the parameter $\pi\xi$, which originates
from the normalization factor of the Coulomb wave function of the
continuous spectrum, the convergence of the $\xi$ expansion is
sufficiently slow. Note also that, since the electron-nucleus
binding for the excited $ns$ electron is weaker than that for the
K-shell electron, the cross section $\sigma^{+}_{ns}$ is suppressed
by the factor of $n^{-3}$ with respect to the cross section
$\sigma^{+}_{1s}$. In the limiting case of a free electron, the
cross section for single photoeffect tends to zero \cite{12}.

{\bf 6.} Let us consider the single ionization of helium-like ions
in the $1sns\,({}^{1,3}\!S)$ states by high-energy photon impact,
which is not followed by excitation of the target. The principal
quantum number $n$ is assumed to be $n \geqslant 2$. For the ground
state, the problem has been studied in Ref.~\cite{27}. The process
can proceed by two different channels. We shall start with
ionization of the K-shell electron. Neglecting the electron-electron
interaction, the amplitude of the process reads
\begin{equation}
\mathcal{A}^{(0)}_{\mathrm{I}}=\mathcal{A}_{1s}\,  ,  \label{eq33}
\end{equation}
where $\mathcal{A}_{1s}$ is given by Eq.~\eqref{eq31}. The total
cross section is just $\sigma_{1s}^+$ given by Eq.~\eqref{eq32},
that is, it keeps the same form for both the singlet and triplet
states. The spin dependence appears, when one takes into account the
electron-electron interaction.

\begin{figure}[h]
\centerline{\includegraphics[scale=0.6]{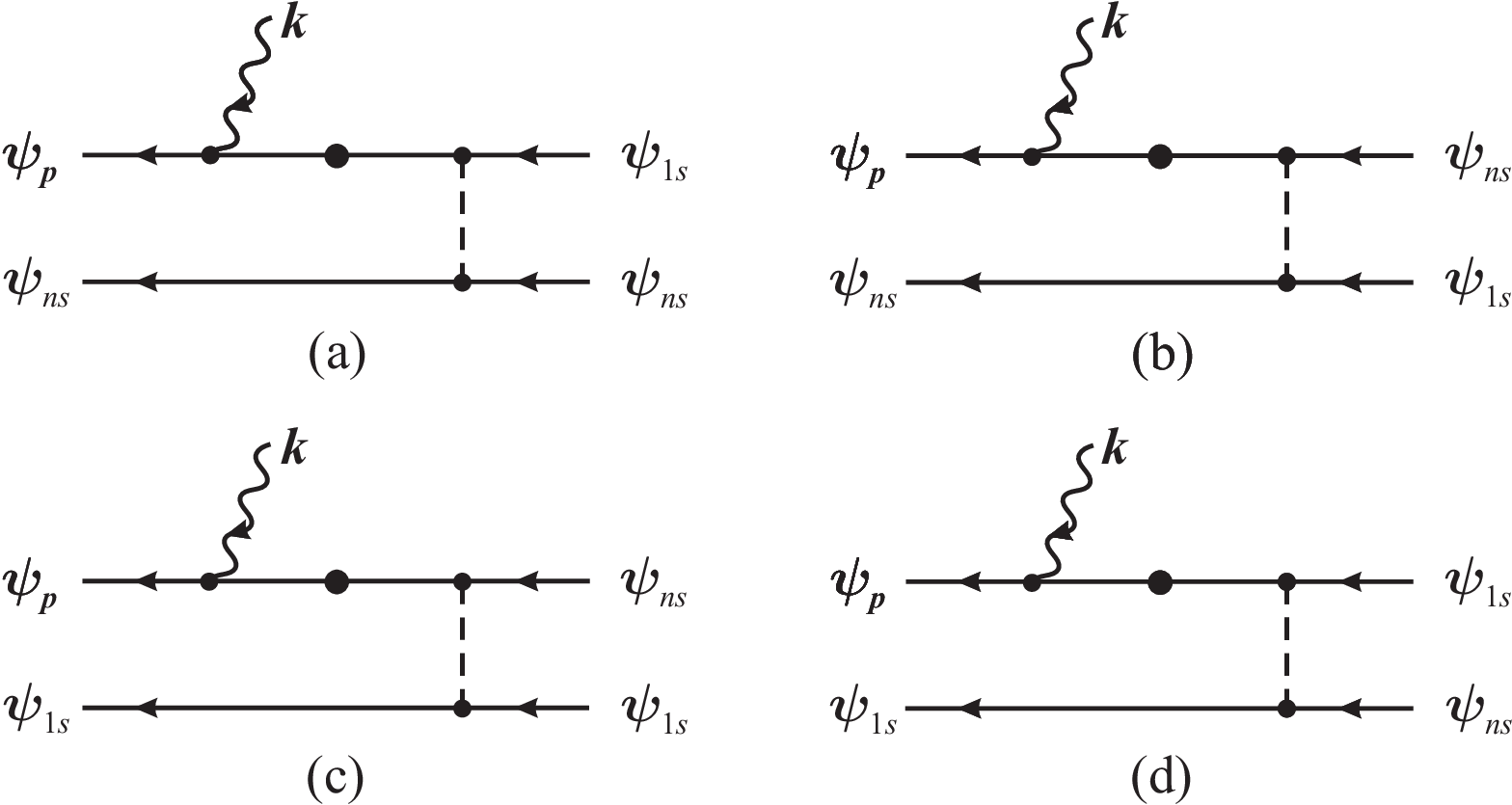}}
\caption{\label{fig3} Feynman diagrams for single ionization of
helium-like ion in the $1sns$ configurations following by the
absorption of a high-energy photon. The lines with a heavy dot
correspond to the reduced Coulomb Green's functions. The diagrams
(a) and (b) describe ionization of the K-shell electron, while
diagrams (c) and (d) describe ionization of the $ns$ electron.}
\end{figure}

In first-order perturbation theory, the amplitude for the first
ionization channel is $\mathcal{A}^{\vphantom{()}}_{\mathrm{I}} =
\mathcal{A}^{(0)}_{\mathrm{I}}+\mathcal{A}^{(1)}_{\mathrm{I}}$.
Within the high-energy asymptotic domain, the dominant contribution
to the correlation correction $\mathcal{A}^{(1)}_{\mathrm{I}}$
arises only from the diagrams depicted in Figs.~\ref{fig3}(a) and
(b), while the other diagrams can be neglected. In explicit terms,
one can write
\begin{eqnarray}
\mathcal{A}^{(1)}_{\mathrm{I}}&=& \mathcal{A}^{(1)}_{\mathrm{a}}
\pm\mathcal{A}^{(1)}_{\mathrm{b}}\,  ,  \label{eq34}\\
\mathcal{A}^{(1)}_{\mathrm{a}} &=& \alpha  N_{\gamma}
N_{1s}\left(\bm{\mathrm e} \cdot \bm{p} \right) \frac{\eta_n}{n^2 }
\frac{1}{(2 \pi i)^2} \oint\limits^{(\eta_n^+)} \! \prod
\limits_{k=1}^{2}\biggl[ dy_k \frac{(y_k+\eta_n)^{n}}{(y_k -
\eta_n)^{n}} \biggr] \nonumber\\
&&\times \Bigl(-\frac{\partial}{\partial \lambda} \Bigr)
\frac{1}{\lambda^{2}}\langle \bm{q}|G_{\mathrm{R}}(E_{1s})
\left(V_{i \eta} - V_{i(\lambda + \eta)}\right) | 0
\rangle_{\left|\vphantom{\bigl(}\, \lambda = y_1 + y_2 \right. }
\,  ,  \label{eq35}\\
\mathcal{A}^{(1)}_{\mathrm{b}} &=& \alpha  N_{\gamma}
N_{1s}\left(\bm{\mathrm e} \cdot \bm{p} \right) \frac{\eta_n}{n^2 }
\frac{1}{(2 \pi i)^2} \oint\limits^{(\eta_n^+)} \! \prod
\limits_{k=1}^{2}\biggl[ dy_k \frac{(y_k+\eta_n)^{n}}{(y_k -
\eta_n)^{n}}\biggr] \nonumber\\
&&\times \Bigl(-\frac{\partial}{\partial \lambda} \Bigr)
\frac{1}{\lambda^{2}}\langle \bm{q}|G_{\mathrm{R}}(E_{1s})
\left(V_{i y_2} - V_{i(\lambda + y_2)}\right) | 0
\rangle_{\left|\vphantom{\bigl(}\, \lambda = y_1 + \eta\right.}\, .
\label{eq36}
\end{eqnarray}
Here again the momentum transfer is $\bm{q}=\bm{p} - \bm{k} \simeq
\bm{p}$. In the derivation, we have used the Born approximation for
the wave function of the ejected high-energy electron. In
Eq.~\eqref{eq34}, the plus and minus signs correspond to the singlet
and triplet states, respectively. For $q \gg \eta$, the matrix
element involving the reduced Coulomb Green's function
$G_{\mathrm{R}}(E_{1s})$ was evaluated in the work \cite{27}.

The correlation corrections for the amplitude can be cast into the
following form
\begin{equation}
\mathcal{A}^{(1)}_{\mathrm{I}}= N_{\gamma} N_{1s}  \dfrac{4 \pi
\eta}{Z q^4} \left(\bm{\mathrm e} \cdot \bm{p} \right) a^{\pm}_1 \,
. \label{eq37}
\end{equation}
The coefficients $a^{\pm}_1$, which correspond to the singlet and
triplet states, are presented in Tables \ref{table6} and
\ref{table7}. In the limit $n\gg 1$, the product $n^3 a^{+}_1$ tends
to the asymptotic constant $-1.031$, while $n^3 a^{-}_1$ approaches
the value $0.087$. Employing Eqs.~\eqref{eq33} and \eqref{eq37}
yields the total amplitude for the single K-shell photoeffect
\begin{equation}
\mathcal{A}^{\vphantom{()}}_{\mathrm{I}}=\mathcal{A}^{(0)}_{\mathrm{I}}+
\mathcal{A}^{(1)}_{\mathrm{I}}=
\mathcal{A}^{(0)}_{\mathrm{I}}\Bigl(1 +  \dfrac{a^{\pm}_1}{2
Z}\Bigr) \,   ,
\end{equation}
which takes into account the electron correlations. The ionization
cross section reads
\begin{equation}
\sigma^+_{\mathrm{I}} = \sigma_{1s}^+ \left( 1 + a^{\pm}_1
Z^{-1}\right) \,  , \label{eq39}
\end{equation}
where $\sigma^+_{1s}$ is given by Eq.~\eqref{eq32}. As it is seen,
account of the electron-electron interaction gives rise to the
significant dependence of ionization cross sections on the spin
multiplicity of atomic states. Indeed, the characteristic orbits of
the $1s$ and $ns$ electrons are somewhat different. However, in the
singlet state, both electrons are allowed to be at the same spatial
point, while, in the triplet state, it is forbidden by the Pauli
principle. Accordingly, in the $n\,{}^{1}\!S$ state, the account of
the dominant correlation corrections attenuates the binding of the
K-shell electron with the nucleus. As a result, the cross section
decreases in comparison with the single-particle prediction
\eqref{eq32}. In the $n\,{}^{3}\!S$ state, the correlation
corrections amplify the electron-nucleus binding for the K-shell
electron, what makes the ionization cross section
$\sigma^+_{\mathrm{I}}$ even a bit larger than $\sigma_{1s}^+$.

Another channel for the single photoeffect on helium-like ions in
the $1sns\,({}^{1,3}\!S)$ states is ionization of the $ns$ electron.
To leading order, the amplitude of the process reads
\begin{equation}
\mathcal{A}^{(0)}_{\mathrm{II}}=\mathcal{A}_{ns}\,  ,  \label{eq40}
\end{equation}
where $\mathcal{A}_{ns}$ is given by Eq.~\eqref{eq31}. The total
cross section coincides with the formula \eqref{eq32}, being
independent of the spin multiplicity of the wave functions.

To first order of the perturbation theory with respect to the
electron-electron interaction, one needs to take into account the
diagrams drawn in Figs.~\ref{fig3}(c) and (d). The corresponding
contributions can be represented as
\begin{eqnarray}
\mathcal{A}^{(1)}_{\mathrm{c}} &=& 2 \alpha \, \eta^2 N_{\gamma}
N_{ns}\left(\bm{\mathrm e} \cdot \bm{p} \right)  \frac{1}{2 \pi i}
\oint\limits^{(\eta_n^+)} \! dy \frac{(y+\eta_n)^{n}}{(y -
\eta_n)^{n}} \nonumber\\
&&\times \Bigl(-\frac{\partial}{\partial \lambda} \Bigr)
\frac{1}{\lambda^{2}}\langle \bm{q}|G_{\mathrm{R}}(E_{ns})
\left(V_{i y} - V_{i(\lambda + y)}\right) | 0
\rangle_{\left|\vphantom{\bigl(}\, \lambda = 2 \eta\right. }
\,  ,  \label{eq41}\\
\mathcal{A}^{(1)}_{\mathrm{d}} &=& 2 \alpha \, \eta^2 N_{\gamma}
N_{ns}\left(\bm{\mathrm e} \cdot \bm{p} \right)  \frac{1}{2 \pi i}
\oint\limits^{(\eta_n^+)} \! dy \frac{(y+\eta_n)^{n}}{(y -
\eta_n)^{n}} \nonumber\\
&&\times \Bigl(-\frac{\partial}{\partial \lambda} \Bigr)
\frac{1}{\lambda^{2}}\langle \bm{q}|G_{\mathrm{R}}(E_{ns})
\left(V_{i \eta} - V_{i(\lambda + \eta)}\right) | 0
\rangle_{\left|\vphantom{\bigl(}\, \lambda = y+ \eta\right. }\, .
\label{eq42}
\end{eqnarray}
Here the matrix elements should be evaluated with the reduced
Green's function $G_{\mathrm{R}}(E_{ns})$ at the energy point
$E_{ns}=-\eta^{2}_{n}/(2m)$, which is related to the usual
non-relativistic Green's function $G(E)$ via
\begin{equation}
G_{\mathrm{R}}(E_{ns}) = \lim_{E \to E_{ns}} \biggl\{ G(E) - \frac{
|\psi_{ns} \rangle  \langle \psi_{ns}| }{E - E_{ns}} \biggr\}   \, .
\label{eq43}                                            
\end{equation}
For $q \gg \eta$, the Coulomb matrix element has the following
integral representation \cite{19}
\begin{equation}
\langle \bm{q} |G(E) V_{i\lambda}| 0 \rangle \simeq 2^5 \pi m
\eta\frac{ip_1}{q^4} \int_0^1 \frac{ t^{-i\zeta } \, dt}{[\lambda -
ip_1 - (\lambda+ip_1)t]^2} \,  , \label{eq44}
\end{equation}
where $\zeta= \eta/p_1$ and $p_1=\sqrt{2 m E}$ is the intermediate
momentum. Performing the analytical continuation of Eq.~\eqref{eq44}
and canceling the pole terms according to the definition
\eqref{eq43} yield
\begin{eqnarray}
\langle \bm{q} |G_{\mathrm{R}}(E_{ns}) V_{i\lambda}| 0 \rangle &=&
2^5 \pi m \frac{\eta^2}{q^4} \biggl\{ P_n(\lambda) + \biggr. \nonumber\\
&&+ (n+1) (\lambda - \eta_{n})^{n} \biggl. \int_0^1 \frac{ \ln t \,
dt}{[\lambda(1-t) + \eta_{n}(1+t)]^{n+2}} \biggr\}\, . \label{eq45}
\end{eqnarray}
The explicit expressions of the function $P_n(\lambda)$ for
particular values of $n\leqslant 9$ are given in Table \ref{table8}.

In the case of ionization of the $ns$ electron, it is convenient to
introduce the correlation coefficients $b^{\pm}_1$ according to
\begin{equation}
\mathcal{A}^{(1)}_{\mathrm{II}}= \mathcal{A}^{(1)}_{\mathrm{c}}
\pm\mathcal{A}^{(1)}_{\mathrm{d}} =N_{\gamma} N_{ns}  \dfrac{4 \pi
\eta}{Z q^4} \left(\bm{\mathrm e} \cdot \bm{p} \right) b^{\pm}_1 \,
, \label{eq46}
\end{equation}
where plus and minus correspond to the singlet and triplet states,
respectively. The coefficients $b^{\pm}_1$ are evaluated in Tables
\ref{table9} and \ref{table10}. In the asymptotic limit $n\gg 1$,
$b^{+}_1$ tends to the constant $-1.47$, while $b^{-}_1$ approaches
to the value $-2.44$. The total amplitude for ionization of the $ns$
electron is $\mathcal{A}^{\vphantom{()}}_{\mathrm{II}} =
\mathcal{A}^{(0)}_{\mathrm{II}} + \mathcal{A}^{(1)}_{\mathrm{II}}$.
The partial cross section for the second channel reads
\begin{equation}
\sigma^+_{\mathrm{II}} = \sigma_{ns}^+ \left( 1 + b^{\pm}_1
Z^{-1}\right) \,  , \label{eq47}
\end{equation}
where $\sigma^+_{ns}$ is given by Eq.~\eqref{eq32}. As it is seen,
account of the electron-electron interaction turns out to be crucial
for describing the high-energy photoeffect on the $ns$ electron.
Especially, it concerns the triplet states, where the screening
corrections $b^{-}_1$ are very large. Due to the Pauli exclusion
principle, the transfer of a large momentum from the nucleus to the
$ns$ electron appears to be hardly probable. Accordingly, the
ionization cross section strongly decreases in comparison with the
single-particle approximation \eqref{eq32}. Note that, for a neutral
helium atom in the $n\,{}^{3}\!S$ state, Eq.~\eqref{eq47} predicts
negative values for $\sigma^+_{\mathrm{II}}$, if $n\geqslant 3$. In
this case, one needs to take into account higher-order correlation
corrections, which have not been considered in the present paper.
For two-electron ions with $Z \geqslant 3$, the cross section
\eqref{eq47} is always positive.

Taking into account both ionization channels, the total cross
section for the single photoeffect on helium-like ions in the
$1sns\, ({}^{1,3}\!S)$ states reads
\begin{eqnarray}
\sigma^+ &=&\sigma^+_{\mathrm{I}}+\sigma^+_{\mathrm{II}}
\label{eq48}\\
&=& \sigma_{1s}^+ \left( 1 + a^{\pm}_1 Z^{-1}\right) + \sigma_{ns}^+
\left( 1 + b^{\pm}_1 Z^{-1}\right) \,  . \label{eq49}
\end{eqnarray}
In view of the relation \eqref{eq32}, Eq.~\eqref{eq49} can be also
cast into the following form
\begin{equation}
\sigma^+ =\sigma_{1s}^+ \bigl\{ 1 + n^{-3} + \left(a^{\pm}_1 +
n^{-3} b^{\pm}_1\right) Z^{-1}\bigl\} \, . \label{eq50}
\end{equation}
The coefficients $a^{\pm}_1$ and $b^{\pm}_1$ describe the dominant
contribution to the correlation effect at high photon energies.

Although the expressions \eqref{eq49} and \eqref{eq50} have been
derived within the Born approximation, Eq.~\eqref{eq49} can be
employed for a sufficiently wide energy domain, provided the
single-particle cross sections are described by Eq.~\eqref{eq29}
\cite{27,28}. Indeed, the expression \eqref{eq29} keeps the same
form to first order of the perturbation theory, taking into account
the correlation corrections to the binding energy. However, for the
high-energy domain characterized by $\varepsilon_{\gamma}^{-1} \ll
\xi\ll 1$, the binding energy corrections are negligibly small with
respect to the photon energy. Since the function \eqref{eq30}
involves the correct dependence on the parameter $\xi$,
Eq.~\eqref{eq49} turns out to be also correct, neglecting terms of
the order of about $(\xi/Z)^2=(m \alpha/p)^2$ (see also extensive
discussions on this topic in Refs.~\cite{29,30,31,32}). In this
paper, we have omitted some correlation corrections to the
amplitude, which are of about $m \alpha/p$ (in particular, those,
which arise due to the final-state interaction). However, the
neglected terms are purely imaginary and do not contribute to the
ionization cross section.

Concluding, we have investigated the processes of excitation and
ionization of light multicharged ions by impact of high-energy
particles, which proceed with participation of the $ns$ electrons.
The dominant correlation corrections to the energy levels and
photoionization cross sections are calculated analytically within
the framework of non-relativistic perturbation theory. As major
result, the universal scalings for the excitation and ionization
cross sections are deduced.

\acknowledgments

The authors acknowledge financial support from BMBF, DFG, GSI, and
INTAS (Grant no. 06-1000012-8881).

\newpage
\begin{table}
\caption{\label{table1} For the pure $ns^2 \,({}^{1}\!S)$ states of
helium-like ions, the dimensionless coefficients $\epsilon_1$ are
tabulated according to Eq.~\eqref{eq12}.}
\begin{center}
\begin{tabular}{r l l r l l} \hline
\multicolumn{1}{c}{$n$} & \multicolumn{2}{c}{$n^2\epsilon_1$} &
\multicolumn{1}{c}{$n$} & \multicolumn{2}{c}{$n^2\epsilon_1$} \\
\cline{2-3}  \cline{5-6} & \multicolumn{1}{c}{analytical} &
\multicolumn{1}{c}{numerical} &  & \multicolumn{1}{c}{analytical} &
\multicolumn{1}{c}{numerical} \\   \hline
1  & $5/4$  & 1.25  & 9  & $2555635959/2147483648$  &  1.19006 \\
2  & $77/64$ & 1.20313 & 10 & $40886039491/34359738368$ & 1.18994  \\
3  & $153/128$  &  1.19531 & 11 & $81765970991/68719476736$ & 1.18985  \\
4  & $19541/16384$  &  1.19269 & 12 & $10465450564505/8796093022208$  & 1.18978  \\
5  & $39043/32768$  &  1.19150 &  13 & $20929978151879/17592186044416$ & 1.18973  \\
6  & $624353/524288$ &  1.19086 & 14 & $334867944396713/281474976710656$  & 1.18969  \\
7  & $1248305/1048576$  & 1.19048 & 15 & $669717016529889/562949953421312$ & 1.18966 \\
8  & $1277999141/1073741824$  &  1.19023
& 16 & $5486195355159701189/4611686018427387904$ & 1.18963 \\
\hline
\end{tabular}
\end{center}
\end{table}

\begin{table}
\caption{\label{table2} For the $1sns \,({}^1 \!S)$ terms of
helium-like ions, the dimensionless coefficients $\epsilon_1$ are
tabulated according to Eqs.~\eqref{eq14}--\eqref{eq16}.}
\begin{center}
\begin{tabular}{r l l } \hline
\multicolumn{1}{c}{$n$} & \multicolumn{2}{c}{$n^2\epsilon_1$}  \\
\cline{2-3}  & \multicolumn{1}{c}{analytical} &
\multicolumn{1}{c}{numerical} \\   \hline
2  & $1352/729$ & 1.8546 \\
3  & $31041/16384$  & 1.8946 \\
4  & $18741536/9765625$  & 1.9191 \\
5  & $87737225/45349632$  & 1.9347 \\
6  & $1319346413640/678223072849$  & 1.9453 \\
7  & $137428433691773/70368744177664$  & 1.9530    \\
8  & $294001828174792832/150094635296999121$  & 1.9588    \\
9  & $12270688433717907483/6250000000000000000$  & 1.9633    \\
10 & $160115079305309345784200/81402749386839761113321$  & 1.9669    \\
11 & $19575441723611998332463421/9937105900423855516680192$  & 1.9699    \\
12 & $180937160122853564881284932640/91733330193268616658399616009$ & 1.9724  \\
13 & $7619039446472988471652076553139/3858646740351497994569680683008$ & 1.9745  \\
14 & $8421485561700554490032347126878248/4261134649619646370410919189453125$ & 1.9763 \\
15 & $168262776200010701611816602447746396925/
85070591730234615865843651857942052864$ & 1.9779 \\
\hline
\end{tabular}
\end{center}
\end{table}

\begin{table}
\caption{\label{table3} For the $1sns\, ({}^3\!S)$ terms of
helium-like ions, the dimensionless coefficients $\epsilon_1$ are
tabulated according to Eqs.~\eqref{eq14}--\eqref{eq16}.}
\begin{center}
\begin{tabular}{r l l } \hline
\multicolumn{1}{c}{$n$} & \multicolumn{2}{c}{$n^2\epsilon_1$} \\
\cline{2-3}  & \multicolumn{1}{c}{analytical} &
\multicolumn{1}{c}{numerical} \\   \hline
2  & $1096/729$  & 1.5034   \\
3  & $27639/16384$ & 1.6869   \\
4  & $17279264/9765625$  & 1.7694   \\
5  & $41200175/22674816$  & 1.8170   \\
6  & $1253464471368/678223072849$  & 1.8482    \\
7  & $131602743884003/70368744177664$  & 1.8702  \\
8  & $283169102583793792/150094635296999121$ & 1.8866   \\
9  & $1483842517395327999/781250000000000000$  & 1.8993  \\
10 & $155435328837703659864200/81402749386839761113321$  & 1.9095  \\
11 & $19056793993685438861087899/9937105900423855516680192$  & 1.9177  \\
12 & $176552744806540188413776122912/91733330193268616658399616009$  & 1.9246 \\
13 & $3724467497547548854690865614181/1929323370175748997284840341504$ & 1.9304 \\
14 & $8247164359458263840860888961901352/4261134649619646370410919189453125$ & 1.9354 \\
15 & $165016222122028120626234494334313260675/
85070591730234615865843651857942052864$ & 1.9398  \\
\hline
\end{tabular}
\end{center}
\end{table}

\begin{table}
\caption{\label{table4} For various values of the principal quantum
number $n$, the dimensionless coefficients $\varkappa_n$ are
calculated according to Eq.~\eqref{eq24}.}
\begin{center}
\begin{tabular}{r l l } \hline
\multicolumn{1}{c}{$n$} & \multicolumn{2}{c}{$\varkappa_n$} \\
\cline{2-3} & \multicolumn{1}{c}{analytical} &
\multicolumn{1}{c}{numerical}\\   \hline
2  & $1048576/295245$  &  3.5515 \\
3  & $85293/35840$  & 2.3798  \\
4  & $806380109824/384521484375$   & 2.0971  \\
5  & $2815390625/1420541793$  & 1.9819   \\
6  & $2238980830283169792/1164315722505971035$ &  1.9230  \\
7  & $835179111639733/442209832796160$  &  1.8886   \\
8  & $156418330091764719082799104/83789232212292561749679885$  &
1.8668  \\
9  & $10197593930567041902351/5506157875061035156250$  & 1.8520  \\
10 &
$580608990927250169145671680000000/315279866676074279955576140196657$
& 1.8416  \\
11 & $28550108411751156734369151319/15568186626467172764709027840$ &
1.8339  \\
12 & $9784817930963041730702096023760230416384/
5352585600085746305433457063492098280775$ & 1.8280  \\
\hline
\end{tabular}
\end{center}
\end{table}

\begin{table}
\caption{\label{table5} For various values of the principal quantum
number $n$, the dimensionless functions $f_n(\xi)$ are tabulated
according to Eqs.~\eqref{eq26} and \eqref{eq28}.}
\begin{center}
\begin{tabular}{r l l } \hline
\multicolumn{1}{c}{$n$}& \  & \multicolumn{1}{c}{$f_n(\xi)$} \\
\hline
1  &&  $2/ \left(1 + {\xi }^2\right)$  \\
2  &&  $32$  \\
3  &&  $54 \left( 27 + 7 {\xi }^2 \right)$  \\
4  &&  $\left(512/3\right)\left( 768 + 288 {\xi }^2 + 23 {\xi }^4 \right) $ \\
5  &&  $\left(1250 /3\right)\left( 46875 + 20625 {\xi }^2 + 2545 {\xi }^4 +
91 {\xi }^6 \right) $  \\
6  &&  $\left(864/5\right)  \left( 25194240 + 12130560 {\xi }^2 +
1827360
{\xi}^4 + 105552 {\xi }^6 + 2023 {\xi }^8   \right) $  \\
7  && $\left(4802/45\right)\left( 12711386205 + 6485401125 {\xi }^2
+ 1097665170 {\xi }^4 + 79704282 {\xi }^6 +
      2547265 {\xi }^8 + 29233 {\xi }^{10} \right) $  \\
8  && $\left(8192/315\right)\left( 21646635171840 + 11499774935040
{\xi }^2 + 2101597962240 {\xi }^4 +
      175037743104 {\xi }^6 + \right.$ \\
   &&  $\left.+ 7190401024 {\xi }^8 + 140890496 {\xi }^{10} +
   1044871 {\xi }^{12} \right)$  \\
9  &&  $\left(4374/35\right) \left( 2402063207770905 +
1314709492319055{\xi }^2 + 253715866938765{\xi }^4 +
      23169940624971 {\xi }^6 + \right.$  \\
    &&  $\left.+ 1109810543211 {\xi }^8 + 28441442925 {\xi }^{10}+
    366767919 {\xi }^{12} + 1859129 {\xi }^{14} \right) $  \\
\hline
\end{tabular}
\end{center}
\end{table}

\begin{table}
\caption{\label{table6} For the $1sns \,({}^1\!S)$ terms of
helium-like ions, the dimensionless coefficients $a^{+}_1$ are
tabulated according to Eqs.~\eqref{eq34}--\eqref{eq37}.}
\begin{center}
\begin{tabular}{r l l } \hline
\multicolumn{1}{c}{$n$} & \multicolumn{2}{c}{$n^3 a^{+}_1$} \\
\cline{2-3} & \multicolumn{1}{c}{analytical} &
\multicolumn{1}{c}{numerical}\\   \hline
2  &  $\left(8/2187\right)\left( -1679 + 3360 \ln (3/2) \right)$ & $-1.1582$  \\
3  &  $\left(81/131072\right) \left( -24245 +  78184 \ln (4/3) \right)$ &
$-1.0832$ \\
4  &  $\left(192/48828125\right) \left(-7097629 + 30599840  \ln (5/4)\right)$ &
$-1.0596$ \\
5  & $\left(125/3265173504 \right) \left(- 1171956731 +
6277663800 \ln (6/5)\right)$ &
$-1.0491$ \\
6  & $\left(648/23737807549715 \right) \left(- 2411780852069 +
15397639065760 \ln (7/6)\right)$ & $-1.0435$ \\
7  & $\left(1029 /5629499534213120 \right) \left(- 496846568524997 +
3678208705153360 \ln (8/7)\right)$ &
$-1.0401$ \\
8  & $\left( 512 /47279810118554723115 \right) \left(- 11062929106398468157 +
93112579989061637760  \ln (9/8)\right)$ &
$-1.0379$ \\
9  & $\left(729 /3500000000000000000000 \right) \left(-
733561568265485829193 + 6915165610828616935920 \ln (10/9)\right)$ &
$-1.0365$ \\
\hline
\end{tabular}
\end{center}
\end{table}

\begin{table}
\caption{\label{table7} For the $1sns \,({}^3\!S)$ terms of
helium-like ions, the dimensionless coefficients $a^{-}_1$ are
tabulated according to Eqs.~\eqref{eq34}--\eqref{eq37}.}
\begin{center}
\begin{tabular}{r l l } \hline
\multicolumn{1}{c}{$n$} & \multicolumn{2}{c}{$n^3 a^{-}_1$} \\
\cline{2-3} & \multicolumn{1}{c}{analytical} &
\multicolumn{1}{c}{numerical}\\   \hline
2  & $\left(8/2187\right) \left(-1399 + 3552 \ln (3/2)\right)$ & $0.1507$  \\
3  & $\left(27/131072\right) \left( -67801 +
237576\ln (4/3) \right)$  &  $0.1123$ \\
4  & $\left(64/48828125 \right) \left(- 20509663 + 92256480  \ln (5/4) \right)$
&  $0.1006$ \\
5  & $\left(125/1632586752 \right) \left(- 572430899 +
3146517000 \ln (6/5) \right)$
& $0.0955$ \\
6  & $\left(216/4747561509943 \right) \left(- 1424070240731 +
9251393817120 \ln (7/6) \right)$
& $0.0927$ \\
7  & $\left( 2401/ 5629499534213120\right) \left(- 210463375781159 +
1577733920822640 \ln (8/7) \right)$ & $0.0911$ \\
8  & $\left( 512/ 6754258588364960445\right) \left(- 1566435008617827011 +
13309413901475690880\ln (9/8) \right)$ & $0.0901$ \\
9  & $\left( 729/ 437500000000000000000 \right) \left(-
91055968618569041179 + 864741335682205139760
\ln (10/9) \right)$ & $0.0893$ \\
\hline
\end{tabular}
\end{center}
\end{table}

\begin{table}
\caption{\label{table8} For various values of the principal quantum
number $n$, the functions $P_n(\lambda)$ are tabulated according to
Eq.~\eqref{eq45}.}
\begin{center}
\begin{tabular}{r l l } \hline
\multicolumn{1}{c}{$n$}& \  & \multicolumn{1}{c}{$P_n(\lambda)$} \\
\hline 1 && $\dfrac{2 \eta}{\left( \eta + \lambda  \right)^3} -
\dfrac{5} {2 \left( \eta + \lambda  \right)^2}$
$\frac{\vphantom{\Bigl(}}{\vphantom{\Bigl(}}$\\
2  &&  $\dfrac{1}{8 \eta_2^2} - \dfrac{3 \lambda^2}{\left( \lambda
+ {\eta_2} \right)^4} -
  \dfrac\lambda{2 \left( \lambda  + {\eta_2} \right)^3} +
  \dfrac{5}{4 \left( \lambda  + {\eta_2} \right)^2}$
  $\frac{\vphantom{\bigl(}}{\vphantom{\Bigl(}}$\\
3  &&  $\dfrac\lambda{24 \eta_3^3} - \dfrac{16 \lambda^3}{3 \left(
\lambda  + {\eta_3} \right)^5} +
  \dfrac{14 \lambda }{3 \left( \lambda  + {\eta_3} \right)^3} -
  \dfrac{5}{3 \left( \lambda  + {\eta_3} \right)^2}$
  $\frac{\vphantom{\bigl(}}{\vphantom{\Bigl(}}$\\
4  &&  $\dfrac{\lambda^2}{64 \eta_4^4} - \dfrac\lambda{48 \eta_4^3}
+ \dfrac{5}{192 \eta_4^2} -
  \dfrac{10 \lambda^4}{\left( \lambda  + \eta_4 \right)^6} +
  \dfrac{7 \lambda^3}{3 \left( \lambda  + \eta_4 \right)^5} +
  \dfrac{13 \lambda^2}{\left( \lambda  + \eta_4 \right)^4} -
  \dfrac{39 \lambda }{4 \left( \lambda  + \eta_4 \right)^3} +
  \dfrac{47}{24  \left( \lambda  + \eta_4 \right)^2}$
  $\frac{\vphantom{\bigl(}}{\vphantom{\Bigl(}}$\\
5  &&  $\dfrac{\lambda^3}{160 \eta_5^5} - \dfrac{\lambda^2}{64
\eta_5^4} +  \dfrac\lambda{60 \eta_5^3} + \dfrac{1}{192 \eta_5^2}-
\dfrac{96 \lambda^5}{5 \left( \lambda + \eta_5\right)^7} +
  \dfrac{148 \lambda^4}{15 \left( \lambda  + \eta_5 \right)^6} +
  \dfrac{472 \lambda^3}{15 \left( \lambda  + \eta_5 \right)^5} -
  \dfrac{38 \lambda^2}{\left( \lambda  + \eta_5 \right)^4} +
  \dfrac{232 \lambda }{15 \left( \lambda  + \eta_5 \right)^3} -$
  $\frac{\vphantom{\bigl(}}{\vphantom{\bigl(}}$\\
  && $-  \dfrac{131}{60 \left( \lambda  + \eta_5 \right)^2}$
  $\frac{\vphantom{\bigl(}}{\vphantom{\Bigl(}}$\\
6  && $\dfrac{\lambda^4}{384 \eta_6^6} - \dfrac{3 \lambda^3}{320
\eta_6^5} +
  \dfrac{13 \lambda^2}{960 \eta_6^4} - \dfrac{7 \lambda }{960 \eta_6^3} +
  \dfrac{17}{1920 \eta_6^2} - \dfrac{112 \lambda^6}{3 \left(
  \lambda  + \eta_6 \right)^8} +
  \dfrac{464 \lambda^5}{15 \left( \lambda  + {\eta_6} \right)^7} +
  \dfrac{208 \lambda^4}{3 \left( \lambda  + {\eta_6} \right)^6} -
  \dfrac{368 \lambda^3}{3 \left( \lambda  + {\eta_6} \right)^5}
  +$ $\frac{\vphantom{\bigl(}}{\vphantom{\bigl(}}$\\
  && $+ \dfrac{229 \lambda^2}{3 \left( \lambda  + {\eta_6} \right)^4} -
  \dfrac{65 \lambda }{3 \left( \lambda  + {\eta_6} \right)^3} +
  \dfrac{71}{30 \left( \lambda  + {\eta_6} \right)^2}$
  $\frac{\vphantom{\bigl(}}{\vphantom{\Bigl(}}$\\
7  && $\dfrac{\lambda^5}{896 \eta_7^7} - \dfrac{\lambda^4}{192
\eta_7^6} +
  \dfrac{67 \lambda^3}{6720 \eta_7^5} - \dfrac{3 \lambda^2}{320 \eta_7^4} +
  \dfrac{71 \lambda }{13440 \eta_7^3} + \dfrac{1}{240 \eta_7^2} -
  \dfrac{512 \lambda^7}{7 \left( \lambda  + {\eta_7} \right)^9} +
  \dfrac{2992 \lambda^6}{35 \left( \lambda  + {\eta_7} \right)^8} +
  \dfrac{4944 \lambda^5}{35 \left( \lambda  + {\eta_7} \right)^7}
  -$   $\frac{\vphantom{\bigl(}}{\vphantom{\bigl(}}$\\
 && $-  \dfrac{2476 \lambda^4}{7 \left( \lambda  + {\eta_7} \right)^6} +
  \dfrac{2104 \lambda^3}{7 \left( \lambda  + {\eta_7} \right)^5} -
  \dfrac{129 \lambda^2}{\left( \lambda  + \eta_7 \right)^4} +
  \dfrac{989 \lambda }{35 \left( \lambda  + \eta_7 \right)^3} -
  \dfrac{353}{140 \left( \lambda  + {\eta_7} \right)^2}$
   $\frac{\vphantom{\bigl(}}{\vphantom{\Bigl(}}$\\
8  && $\dfrac{\lambda^6}{2048 \eta_8^8} - \dfrac{5 {\lambda}^5}{1792
\eta_8^7} +
  \dfrac{289\lambda^4}{43008 \eta_8^6} - \dfrac{29\lambda^3}{3360\eta_8^5} +
  \dfrac{457\lambda^2}{71680\eta_8^4} - \dfrac{53\lambda }{26880\eta_8^3} +
  \dfrac{919}{215040 \eta_8^2} - \dfrac{144 \lambda^8}{\left(
  \lambda  + \eta_8 \right)^{10}} +
  \dfrac{7704\lambda^7}{35\left( \lambda  + {\eta_8} \right)^9}
  +$ $\frac{\vphantom{\bigl(}}{\vphantom{\bigl(}}$\\
 && $+  \dfrac{1328\lambda^6}{5\left( \lambda  + \eta_8 \right)^8} -
  \dfrac{4722\lambda^5}{5\left( \lambda  + \eta_8 \right)^7} +
  \dfrac{1032\lambda^4}{\left( \lambda  + \eta_8 \right)^6} -
  \dfrac{1193\lambda^3}{2\left( \lambda  + \eta_8 \right)^5} +
  \dfrac{984\lambda^2}{5\left( \lambda  + \eta_8 \right)^4} -
  \dfrac{1407\lambda }{40\left( \lambda  + \eta_8 \right)^3} +
  \dfrac{1487}{560\left( \lambda  + \eta_8 \right)^2}$
  $\frac{\vphantom{\bigl(}}{\vphantom{\Bigl(}}$ \\
9  && $\dfrac{\lambda^7}{4608\eta_9^9} -
\dfrac{3\lambda^6}{2048\eta_9^8}+
  \dfrac{275\lambda^5}{64512\eta_9^7} - \dfrac{99\lambda^4}{14336\eta_9^6} +
  \dfrac{17\lambda^3}{2520\eta_9^5} - \dfrac{279\lambda^2}{71680\eta_9^4} +
  \dfrac{523\lambda }{322560\eta_9^3} + \dfrac{207}{71680\eta_9^2} -
  \dfrac{2560\lambda^9}{9\left( \lambda  + \eta_9 \right)^{11}} +$
  $\frac{\vphantom{\bigl(}}{\vphantom{\bigl(}}$ \\
 && $+  \dfrac{56864\lambda^8}{105\left( \lambda  + \eta_9 \right)^{10}} +
  \dfrac{141952\lambda^7}{315\left( \lambda  + \eta_9 \right)^9} -
  \dfrac{107168 \lambda^6}{45\left( \lambda  + \eta_9 \right)^8} +
  \dfrac{48416 \lambda^5}{15\left( \lambda  + \eta_9 \right)^7} -
  \dfrac{21236 \lambda^4}{9\left( \lambda  + \eta_9 \right)^6} +
  \dfrac{46952 \lambda^3}{45\left( \lambda  + \eta_9 \right)^5} -$
  $\frac{\vphantom{\bigl(}}{\vphantom{\bigl(}}$ \\
 && $-  \dfrac{1402 \lambda^2}{5\left( \lambda  + \eta_9 \right)^4} +
  \dfrac{13348 \lambda }{315\left( \lambda  + \eta_9 \right)^3} -
  \dfrac{6989}{2520\left( \lambda  + \eta_9 \right)^2}$
  $\frac{\vphantom{\bigl(}}{\vphantom{\Bigl(}}$ \\
\hline
\end{tabular}
\end{center}
\end{table}

\begin{table}
\caption{\label{table9} For the $1sns \,({}^1\!S)$ terms of
helium-like ions, the dimensionless coefficients $b^{+}_1$ are
tabulated according to Eqs.~\eqref{eq41}, \eqref{eq42}, and
\eqref{eq46}.}
\begin{center}
\begin{tabular}{r l l } \hline
\multicolumn{1}{c}{$n$} & \multicolumn{2}{c}{$b^{+}_1$} \\
\cline{2-3} & \multicolumn{1}{c}{analytical} &
\multicolumn{1}{c}{numerical}\\   \hline
2  &  $\left(4/2187\right)\left( -772 + 159 \ln 3 \right)$ & $-1.0925$  \\
3  &  $\left(3/131072 \right)\left( -73875 + 13816 \ln 4 \right)$ & $-1.2525$  \\
4  &  $\left(72/48828125 \right)\left( -1248684 + 219365 \ln 5 \right)$ & $-1.3207$  \\
5  &  $\left( 5/3265173504 \right)\left( -1268946065 + 213266808 \ln 6 \right)$ &
$-1.3580$  \\
6  &  $\left( 12/23737807549715 \right)\left( -3996048740304 +
649245311015 \ln 7 \right)$ &
$-1.3814$  \\
7  &  $\left( 7/ 5629499534213120\right)\left( -1674319803898931 +
264724373084400 \ln 8 \right)$ &
$-1.3974$  \\
8  &  $\left( 16/47279810118554723115 \right)\left( -6304966995252184568 +
974522181024852075  \ln 9 \right)$ & $-1.4090$  \\
9  &  $\left( 9/3500000000000000000000 \right)\left( -847067488481686729599 +
128414477117971809520\ln 10 \right)$ & $-1.4178$  \\
\hline
\end{tabular}
\end{center}
\end{table}

\begin{table}
\caption{\label{table10} For the $1sns\,({}^3\!S)$ terms of
helium-like ions, the dimensionless coefficients $b^{-}_1$ are
tabulated according to Eqs.~\eqref{eq41}, \eqref{eq42}, and
\eqref{eq46}.}
\begin{center}
\begin{tabular}{r l l } \hline
\multicolumn{1}{c}{$n$} & \multicolumn{2}{c}{$b^{-}_1$} \\
\cline{2-3} & \multicolumn{1}{c}{analytical} &
\multicolumn{1}{c}{numerical}\\   \hline
2  &  $\left(4/2187\right)\left( -1604 + 543 \ln 3  \right)$ & $-1.8426$  \\
3  &  $\left(3/131072 \right)\left(-146181 + 41032 \ln 4 \right)$ & $-2.0439$  \\
4  &  $\left(8 /48828125 \right)\left( -22143244 + 5629965  \ln 5 \right)$ &
$-2.1434$  \\
5  &  $\left(5 /1632586752 \right)\left( -1254848345 + 298760904 \ln 6 \right)$ &
$-2.2037$  \\
6  &  $\left(12 /4747561509943 \right)\left( -1589311619664 + 360435860155 \ln 7 \right)$ &
$-2.2444$  \\
7  &  $\left(21 /5629499534213120 \right)\left( -1116056979027599 +
243593185902000 \ln 8 \right)$ & $-2.2737$  \\
8  &  $\left( 16/6754258588364960445 \right)\left( -1810645396585286728 +
382953780229600125  \ln 9 \right)$ &
$-2.2959$  \\
9  &  $\left( 9/ 437500000000000000000\right)\left( -213879211569028496997 +
44048190258616320560\ln 10 \right)$ & $-2.3133$  \\
\hline
\end{tabular}
\end{center}
\end{table}

\clearpage


\clearpage

\begin{thebibliography}{99}
\bibitem{13} The use of the same notations for the electron mass and
projection of the orbital angular momentum is not confusing, because
the latter is related with the angular dependence of wave functions
only.
\end{thebibliography}
\end{document}